\documentclass[10pt, conference, letterpaper]{ IEEEtran}

\usepackage{cite}
\usepackage{amsmath,amssymb,amsfonts}
\usepackage{algorithmic}
\usepackage{graphicx}
\usepackage{textcomp}
\usepackage{xcolor}
\usepackage{tikz}

\usepackage{siunitx} 
\usepackage{bm}
\usepackage{comment}
\usepackage{subcaption}
\usepackage{caption}
\usepackage{dblfloatfix} 

\usetikzlibrary{calc}
\def\BibTeX{{\rm B\kern-.05em{\sc i\kern-.025em b}\kern-.08em
    T\kern-.1667em\lower.7ex\hbox{E}\kern-.125emX}}

\begin{document}

\title{Device-Centric ISAC for Exposure Control via Opportunistic Virtual Aperture Sensing}

\author{
\IEEEauthorblockN{Marouan Mizmizi\IEEEauthorrefmark{1}, Zhibin Yu\IEEEauthorrefmark{2}, Guanglong Du\IEEEauthorrefmark{3}, and Umberto Spagnolini\IEEEauthorrefmark{1}}
\IEEEauthorblockA{\IEEEauthorrefmark{1}Dipartimento di Elettronica, Informazione e Bioingegneria, Politecnico di Milano, Milan, Italy}
\IEEEauthorblockA{\IEEEauthorrefmark{2} Huawei Heisenberg Research Center (Munich), Huawei Technologies Duesseldorf GmbH, Germany}
\IEEEauthorblockA{\IEEEauthorrefmark{3}Huawei Technologies Co., Ltd., Beijing, China}
}

\maketitle

\begin{abstract}
Regulatory limits on Maximum Permissible Exposure (MPE) require handheld devices to reduce transmit power when operated near the user's body. Current proximity sensors provide only binary detection, triggering conservative power back-off that degrades link quality. If the device could measure its distance from the body, transmit power could be adjusted proportionally, improving throughput while maintaining compliance.
This paper develops a device-centric integrated sensing and communication (ISAC) method for the device to measure this distance. The uplink communication waveform is exploited for sensing, and the natural motion of the user's hand creates a virtual aperture that provides the angular resolution necessary for localization. Virtual aperture processing requires precise knowledge of the device trajectory, which in this scenario is opportunistic and unknown. One can exploit onboard inertial sensors to estimate the device trajectory; however, the inertial sensors accuracy is not sufficient. To address this, we develop an autofocus algorithm based on extended Kalman filtering that jointly tracks the trajectory and compensates residual errors using phase observations from strong scatterers. The Bayesian Cram\'{e}r--Rao bound for localization is derived under correlated inertial errors. Numerical results at 28~GHz demonstrate centimeter-level accuracy with realistic sensor parameters.

\end{abstract}

\begin{IEEEkeywords}
Integrated sensing and communication, virtual aperture, MPE compliance, near-field localization, 6G.
\end{IEEEkeywords}

%==============================================================================
\section{Introduction}
\label{sec:intro}
%==============================================================================

The exposure of the human body to radiofrequency electromagnetic waves is regulated, in near field, through limits on the \emph{maximum permissible exposure} (MPE).  Regulatory frameworks accordingly prescribe maximum admissible levels, and portable devices are required to demonstrate compliance under worst-case operating conditions, notably when the handset is used in close proximity to the head during voice services~\cite{ICNIRP2020,FCC_47CFR1310}.

Current devices employ proximity sensors to detect the presence of a conductive body to trigger transmit power back-off when the separation falls below approximately 25~mm~\cite{Semtech_SAR}. This mechanism is essentially binary: it provides no estimate of the separation distance and thus enforces conservative operation. The throughput degradation becomes acute in fifth-generation (5G) networks at millimeter-wave (mmWave) frequencies, where the link budget is already constrained by propagation losses~\cite{Rappaport2019}. If the smartphone could \emph{measure} its distance from the head, proportional power control would become feasible.

Integrated Sensing and Communication (ISAC) has emerged as a transformative paradigm for sixth-generation (6G) networks, enabling simultaneous wireless communication and sensing through shared waveforms, spectrum, and hardware~\cite{Liu2022JSAC,10264814}. Prior work has addressed network-centric localization from base-station infrastructure~\cite{10916597,Wymeersch2019NF,Huang2024Tutorial,NearFieldISAC2024}, bistatic vehicular configurations~\cite{Barneto2024Bistatic,Zeng2022FD_ISAC}, and virtual aperture techniques with motion compensation~\cite{Cumming2005,Carrara1995,Wahl1994,Meta2008}. Device-centric ISAC, where a handheld terminal performs monostatic sensing using its own motion, remains largely unexplored.

This paper proposes a device-centric ISAC framework for smartphone MPE compliance based on opportunistic virtual aperture formation. The device exploits its motion during normal operation to synthesize a virtual sensing aperture, enabling accurate localization of the user's body for distance-proportional power control. The main contributions are as follows. First, we formulate the near-field monostatic sensing problem at mmWave frequencies, modeling the signal, antenna configuration, and inertial measurement unit (IMU) based positioning of the device, and compensating for the errors. Second, we derive the Bayesian Cram\'{e}r-Rao Bound for target localization that incorporates aperture position uncertainty through a correlated Gaussian prior, revealing the fundamental accuracy limits imposed by IMU errors. Third, we develop a complete sensing algorithm comprising relative position-based, phase autofocusing, and backprojection imaging, with explicit power control mapping for MPE compliance. Numerical simulation results demonstrate that centimeter-level localization accuracy is achievable at 28~GHz with realistic motion trajectories and sensor specifications.

The remainder of this paper is organized as follows. Section~\ref{sec:system} introduces the system model, including the monostatic signal model and the IMU-based trajectory error characterization. Section~\ref{sec:crb} derives the Bayesian Cram\'{e}r--Rao bound for target localization under correlated aperture uncertainty. Section~\ref{sec:processing} develops the EKF-based autofocus algorithm. Section~\ref{sec:results} presents numerical results validating the bounds and quantifying the EIRP gain. Section~\ref{sec:conclusion} concludes the paper.
%==============================================================================
\section{System Model}
\label{sec:system}
%==============================================================================
\begin{figure}[b]
    \centering
    \includegraphics[width=0.8\linewidth]{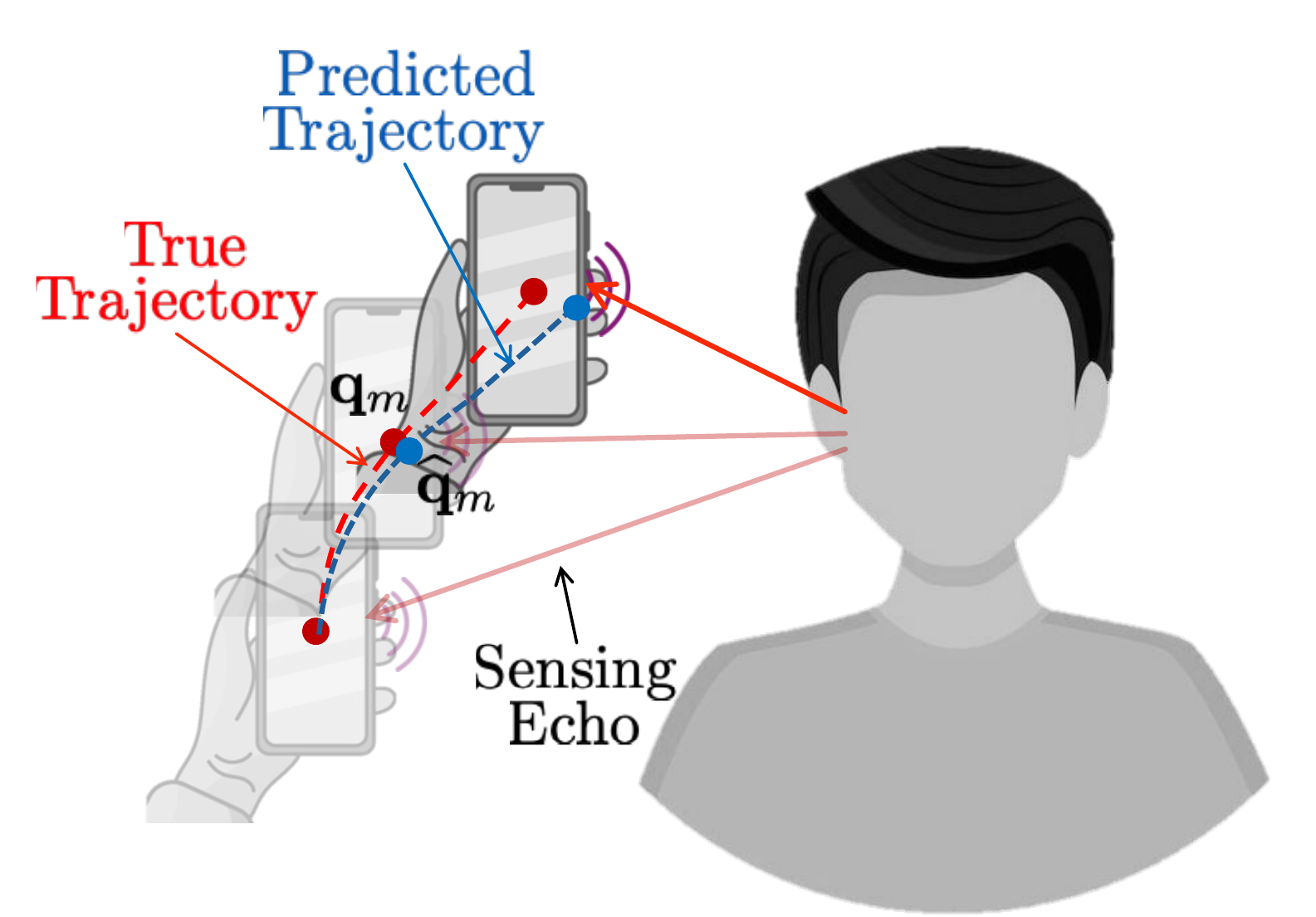}
    \caption{Device-centric ISAC for exposure control: the smartphone exploits natural hand motion to synthesize a virtual aperture, enabling distance measurement for proportional power control. The array phase center follows trajectory $\{\mathbf{q}_m\}$. The IMU provides estimates $\hat{\mathbf{q}}_m$; the error $\boldsymbol{\delta}_m = \hat{\mathbf{q}}_m - \mathbf{q}_m$ must be compensated for coherent synthesis.}
    \label{fig:scenario}
\end{figure}

Fig.~\ref{fig:scenario} illustrates the device-centric sensing geometry considered in this work. A handheld device operates in close proximity to the user’s body and exploits its natural motion to perform monostatic sensing while transmitting uplink communication signals. The sensing task is to localize a small set of dominant reflection points on the body using the communication waveform itself.

The device–body distance exceeds the Fraunhofer distance of the physical antenna array, so that far-field power-density scaling proportional to $r^{-2}$ applies for MPE compliance. At the same time, the same distance lies within the Fresnel region of the motion-synthesized aperture, requiring a near-field formulation with spherical wavefronts for coherent focusing.

Let $\mathbf{q}_m \in \mathbb{R}^3$ denote the position of the array phase center at observation index $m\in\{0,1,\ldots,M-1\}$. The sensing environment is modeled by $Q$ point scatterers at locations $\mathbf{p}_q \in \mathbb{R}^3$ with complex reflectivities $\sigma_q$, representing distinct reflection points on the user’s body. The position of antenna element $n$ at time $m$ is $\mathbf{p}_{n,m}=\mathbf{q}_m+\mathbf{d}_n,$,
where $\mathbf{d}_n$ is the displacement of element $n$ from the array phase center. The corresponding bistatic range is
\begin{equation}
r_{n,q,m}=\|\mathbf{p}_q-\mathbf{p}_{n,m}\|,
\end{equation}
with round-trip delay $\tau_{n,q,m}=2r_{n,q,m}/c$.
The sequence $\{\mathbf{q}_m\}$ defining the virtual aperture is not known exactly. The device relies on an onboard IMU to estimate the phase-center trajectory, yielding position estimates $\{\hat{\mathbf{q}}_m\}$. These estimates are affected by residual errors, $\mathbf{q}_m=\hat{\mathbf{q}}_m-\boldsymbol{\delta}_m$,
where $\boldsymbol{\delta}_m$ denotes the unknown position error at observation $m$.

\subsection{Signal Model}

The transmitted signal is an OFDM waveform with $K$ subcarriers spanning bandwidth $B$ and spacing $\Delta f = B/K$. The $k$-th subcarrier occupies frequency $f_k = f_c + k\Delta f$ for $k = 0, \ldots, K-1$. At observation $m$, the received signal on antenna $n$ comprises self-interference, target echoes, and noise. In the frequency domain, the $k$-th subcarrier sample is
\begin{equation}
Y_{n,m}[k] = \gamma \, S_m[k] + \sum_{q=1}^{Q} \frac{\alpha_q}{r_{n,q,m}^2} \, e^{-j 2\pi f_k \tau_{n,q,m}} S_m[k] + W_{n,m}[k],
\label{eq:rx_full}
\end{equation}
where $S_m[k]$ is the transmitted symbol on subcarrier $k$, $\gamma$ the residual self-interference coefficient, and $W_{n,m}[k] \sim \mathcal{CN}(0, \sigma_w^2)$ is additive noise. The complex amplitude
\begin{equation}
\alpha_q = \sqrt{\frac{P_t G_t G_r \lambda^2 \sigma_{\mathrm{rcs}}}{(4\pi)^3}} \, \chi_q \, e^{j\theta_q}
\label{eq:alpha_q}
\end{equation}
incorporates the reflective cross-section $\sigma_{\mathrm{rcs}}$, antenna gains $G_t$ and $G_r$, cross-polarization coupling $\chi_q$, and random phase $\theta_q$. Cross-polarization between transmit and receive paths provides self-interference suppression; the residual $|\gamma|^2$ is typically below $-30$~dB~\cite{Ericsson_ISAC2024, Zeng2022FD_ISAC}.

Equalization of \eqref{eq:rx_full} and subtraction of the estimated self-interference $\hat{\gamma}$ yields
\begin{equation}
\tilde{Y}_{n,m}[k] = \sum_{q=1}^{Q} \frac{\alpha_q}{r_{n,q,m}^2} \, e^{-j \frac{4\pi}{\lambda} r_{n,q,m}} \cdot e^{-j 2\pi k \Delta f \tau_{n,q,m}} + \tilde{W}_{n,m}[k].
\label{eq:equalized}
\end{equation}
Range compression via inverse DFT yields $z_{n,m}[\ell] = (1/K)\sum_{k} \tilde{Y}_{n,m}[k] e^{j2\pi k\ell/K}$, 
which evaluates to
\begin{equation}
z_{n,m}[\ell] = \sum_{q=1}^{Q} \frac{\alpha_q}{r_{n,q,m}^2} e^{-j\frac{4\pi}{\lambda} r_{n,q,m}} S(\ell - \nu_{n,q,m}) + \tilde{w}_{n,m}[\ell],
\label{eq:z_ell_sinc}
\end{equation}
where $\nu_{n,q,m} = 2Br_{n,q,m}/c$ and $S(\nu) = \sin(\pi\nu)/(K\sin(\pi\nu/K))$ is the Dirichlet kernel with first null at $|\nu|=1$, corresponding to range resolution $\delta_r = c/(2B)$.

\subsection{Trajectory Error Model}

The IMU provides measurements of linear acceleration at each observation instant. The measured acceleration $\tilde{\mathbf{a}}_m = \mathbf{a}_m + \mathbf{b}_a + \mathbf{n}_{a,m}$, where $\mathbf{b}_a$ is the bias and $\mathbf{n}_{a,m} \sim \mathcal{N}(\mathbf{0}, \sigma_a^2 \mathbf{I}_3)$ is measurement noise. Double integration yields position estimates $\hat{\mathbf{q}}_m$ with error $\boldsymbol{\delta}_m = \hat{\mathbf{q}}_m - \mathbf{q}_m$ growing quadratically due to bias~\cite{Sandia2015}.

We model the bias as a random vector $\mathbf{b}_a \sim \mathcal{N}(\mathbf{0}, \sigma_b^2\mathbf{I}_3)$ to capture uncertainty The stacked error vector $\boldsymbol{\delta} = [\boldsymbol{\delta}_1^T, \ldots, \boldsymbol{\delta}_{M-1}^T]^T \in \mathbb{R}^{3(M-1)}$ is then jointly Gaussian:
\begin{equation}
\boldsymbol{\delta} \sim \mathcal{N}(\mathbf{0}, \mathbf{C}_{\boldsymbol{\delta}}),
\label{eq:delta_prior_imu}
\end{equation}
where the covariance factorizes as $\mathbf{C}_{\boldsymbol{\delta}} = \mathbf{C}_t \otimes \mathbf{I}_3$. The temporal correlation matrix $\mathbf{C}_t \in \mathbb{R}^{(M-1)\times(M-1)}$ has elements
\begin{align}
[\mathbf{C}_t]_{mn} &= \frac{\sigma_b^2 T^4}{4} m(m+1)n(n+1)  \notag\\ 
&+\sigma_a^2 T^4 \sum_{i=1}^{\min(m,n)} (m-i+1)(n-i+1),\label{eq:Ct}
\end{align}
reflecting the correlated error growth induced by double integration of bias and noise. The first term, proportional to $\sigma_b^2$, grows as $m^2 n^2$ and dominates at large $m$. The second term, proportional to $\sigma_a^2$, captures the integrated noise.

\section{Bayesian Cram\'{e}r--Rao Bound Analysis}
\label{sec:crb}
%==============================================================================

We derive the Bayesian Cram\'{e}r--Rao Bound (BCRB) for target localization, treating the reflectivity as an unknown nuisance parameter and incorporating prior information on aperture position errors.

\subsection{Vector Observation Model}

From \eqref{eq:z_ell_sinc}, consider a single dominant scatterer and a single antenna (the extension to multiple elements and scatterers affects only the interference structure and is not pursued in the bound). For each aperture index $m$, collect the range-compressed samples in the vector
\begin{equation}
\mathbf{z}_m \triangleq \big[z_m[0],\,z_m[1],\,\ldots,\,z_m[K-1]\big]^T \in \mathbb{C}^{K}.
\end{equation}
Its conditional mean is
\begin{equation}
\boldsymbol{\mu}_m \triangleq \mathbb{E}\{\mathbf{z}_m\}
= \frac{\alpha}{r_m^2}\,e^{-j\kappa r_m}\,\mathbf{s}(\nu_m),
\label{eq:mu_vec}
\end{equation}
where $\kappa \triangleq 4\pi/\lambda$, $\nu_m \triangleq 2Br_m/c$, and $\mathbf{s}(\nu)\in\mathbb{C}^{K}$ has entries
\begin{equation}
[\mathbf{s}(\nu)]_\ell \triangleq S(\ell-\nu), \qquad \ell=0,\ldots,K-1,
\label{eq:s_vec}
\end{equation}
with $S(\cdot)$ is the Dirichlet Kernel. The noise is modeled as $\mathbf{w}_m\sim\mathcal{CN}(\mathbf{0},\sigma_w^2\mathbf{I}_K)$, independent across $m$.

The dependence of $\boldsymbol{\mu}_m$ on $r_m$ enters both through the carrier phase and through the delay index $\nu_m$. Differentiating~\eqref{eq:mu_vec} yields the radial derivative
\begin{equation}
\frac{\partial \boldsymbol{\mu}_m}{\partial r_m}
=
\frac{\alpha}{r_m^2}\,e^{-j\kappa r_m}
\left(
\beta_m\,\mathbf{s}(\nu_m) - \frac{2B}{c}\,\mathbf{s}'(\nu_m)
\right),
\label{eq:dmu_dr}
\end{equation}
where $\beta_m \triangleq -2/r_m - j\kappa$, and $\mathbf{s}'(\nu)$ denotes the derivative with respect to $\nu$,
\begin{equation}
[\mathbf{s}'(\nu)]_\ell \triangleq \frac{\partial}{\partial \nu} S(\ell-\nu) = -S'(\ell-\nu).
\label{eq:sprime}
\end{equation}
It is convenient to define the (dimensionless) \emph{radial sensitivity factor}
\begin{equation}
\mathcal{I}_m
\triangleq
\frac{ \left\| \beta_m\,\mathbf{s}(\nu_m) - \frac{2B}{c}\,\mathbf{s}'(\nu_m) \right\|^2 }
{ \left\| \mathbf{s}(\nu_m) \right\|^2 }.
\label{eq:Im_def}
\end{equation}
In the phase-dominated regime, $\mathcal{I}_m \approx \kappa^2$; the additional term involving $\mathbf{s}'(\nu_m)$ accounts for the information carried by the range-compression envelope.

\begin{figure*}[h!]
    \centering
    % ---- Tune these if needed
    \setlength{\tabcolsep}{1.5pt} % horizontal spacing between subfigures
    \renewcommand{\arraystretch}{1.0}

    \begin{tabular}{cccc}
        \begin{subfigure}[t]{0.2\textwidth}
            \centering
            \includegraphics[width=\linewidth]{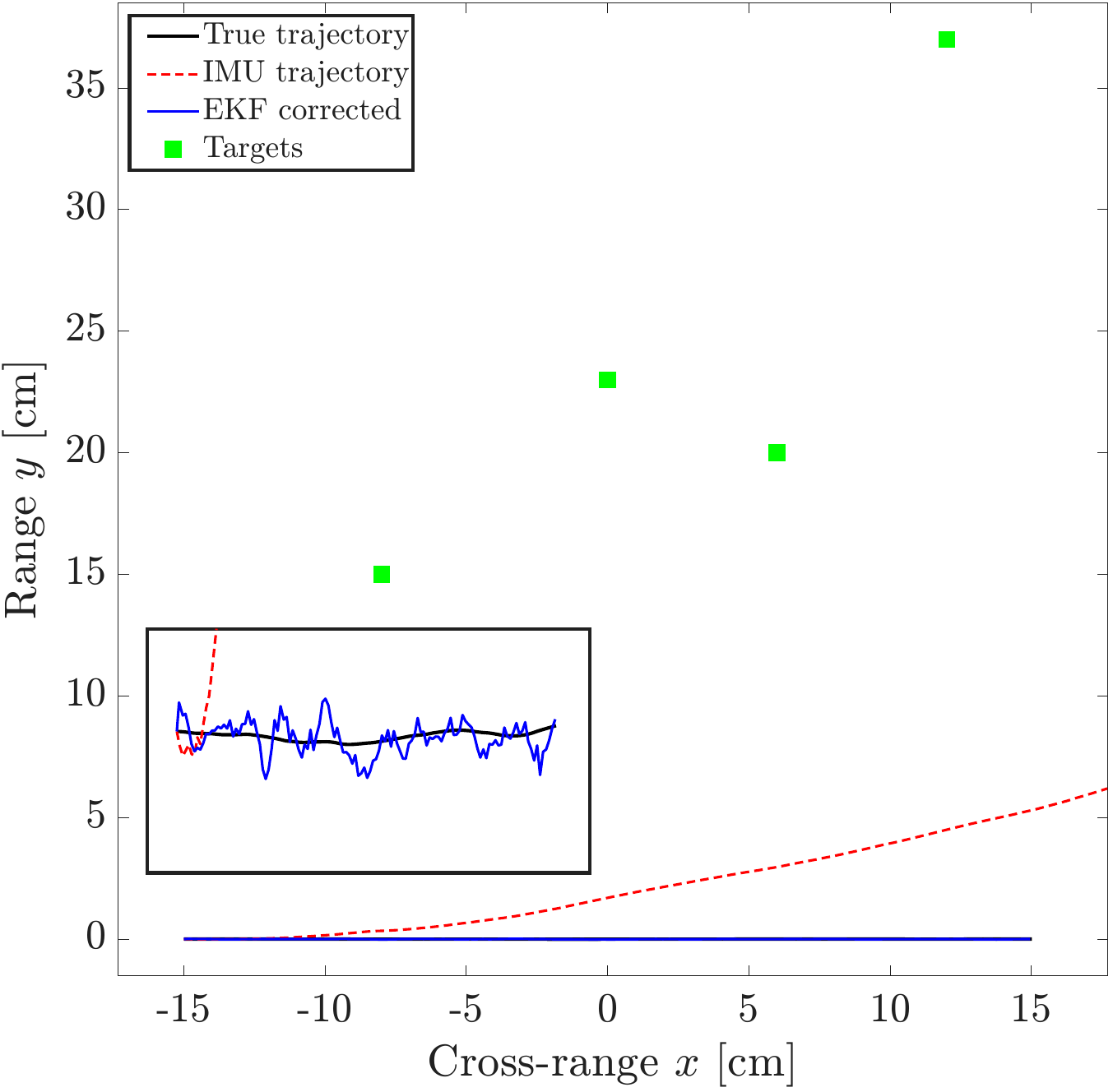}
            \caption{Trajectory: true vs.\ IMU vs.\ EKF.}
            \label{fig:autofocus_traj}
        \end{subfigure} &
        \begin{subfigure}[t]{0.25\textwidth}
            \centering
            \includegraphics[width=\linewidth]{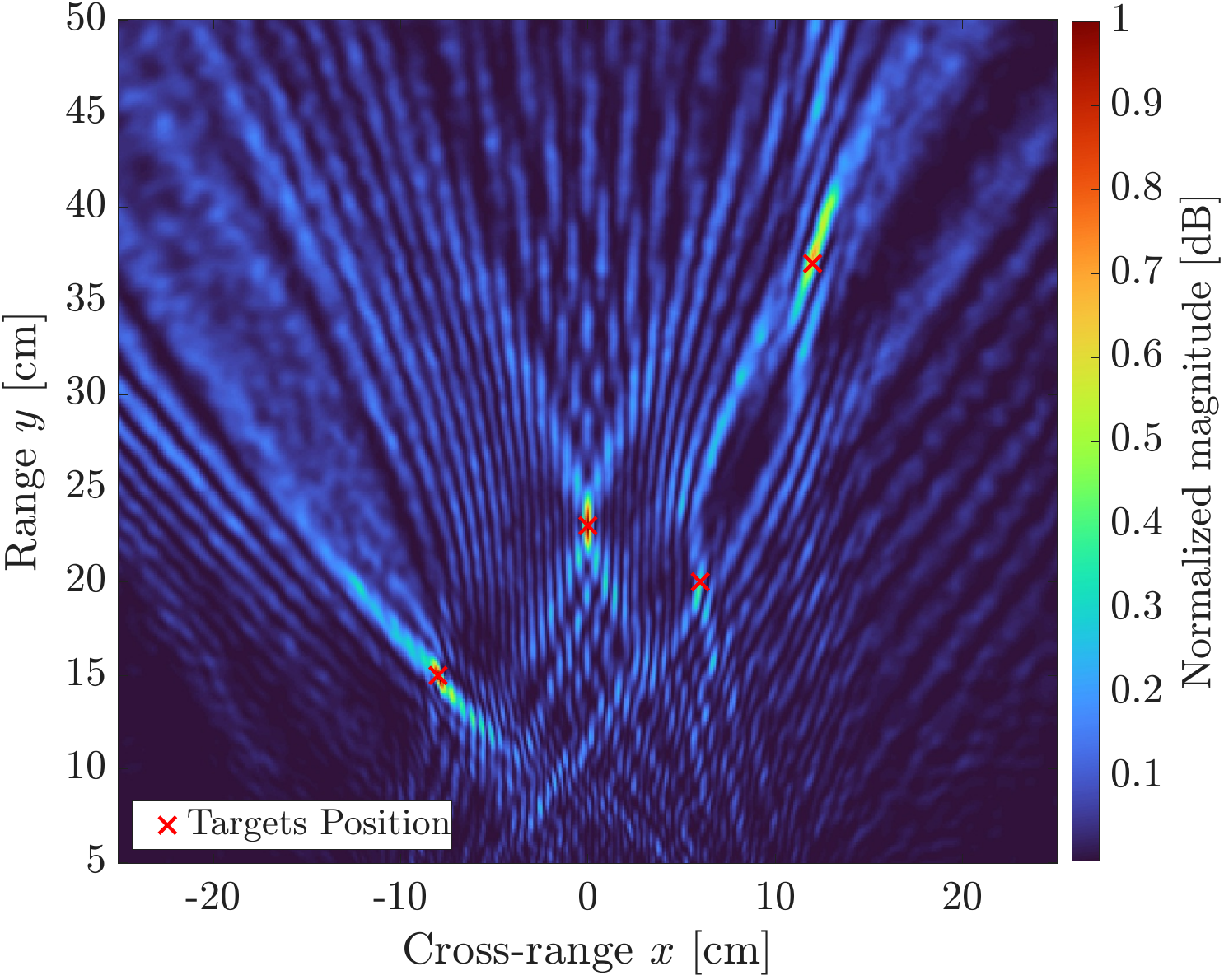}
            \caption{Oracle focusing (true trajectory).}
            \label{fig:autofocus_oracle}
        \end{subfigure} &
        \begin{subfigure}[t]{0.25\textwidth}
            \centering
            \includegraphics[width=\linewidth]{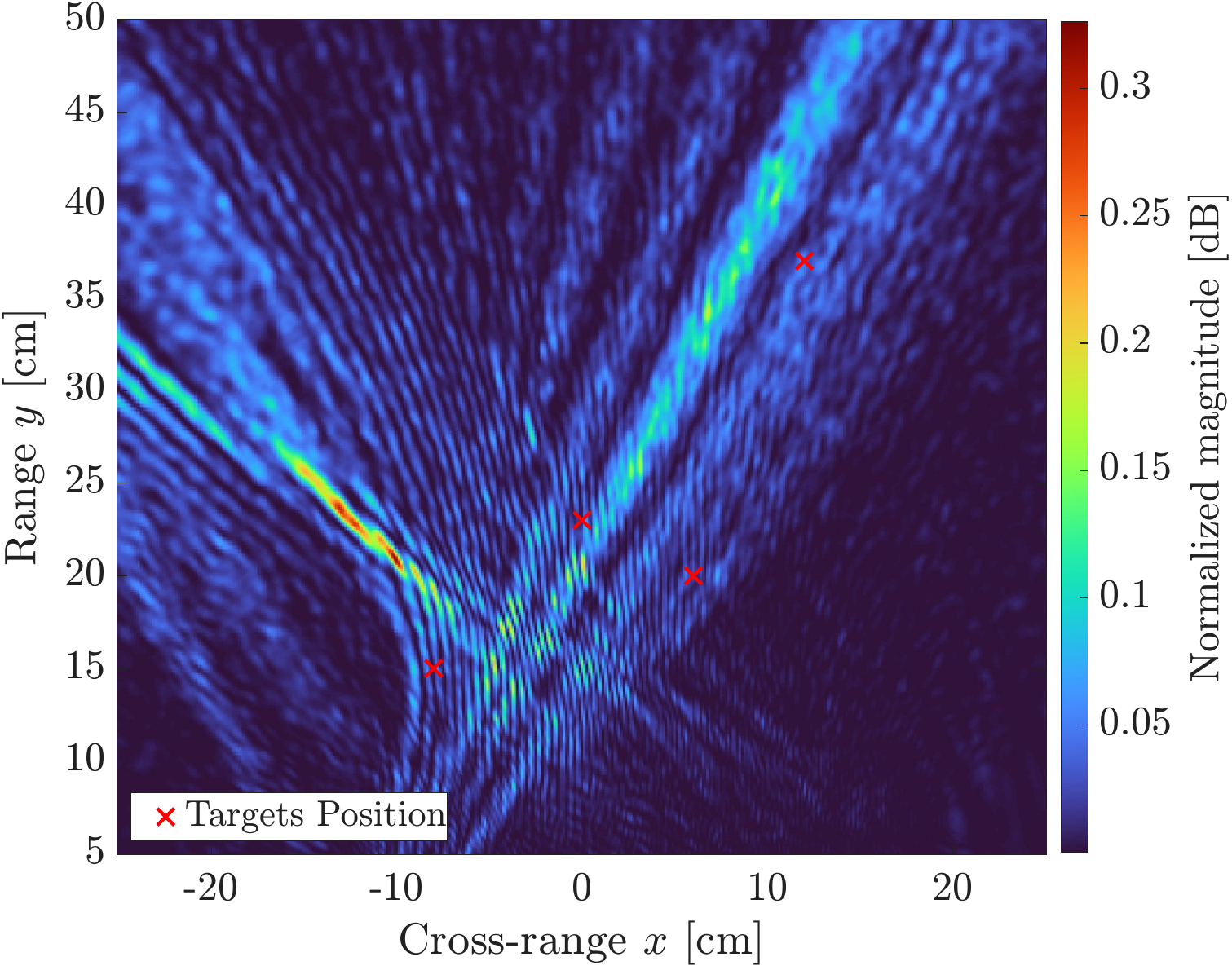}
            \caption{Focusing with IMU trajectory.}
            \label{fig:autofocus_imu}
        \end{subfigure} &
        \begin{subfigure}[t]{0.25\textwidth}
            \centering
            \includegraphics[width=\linewidth]{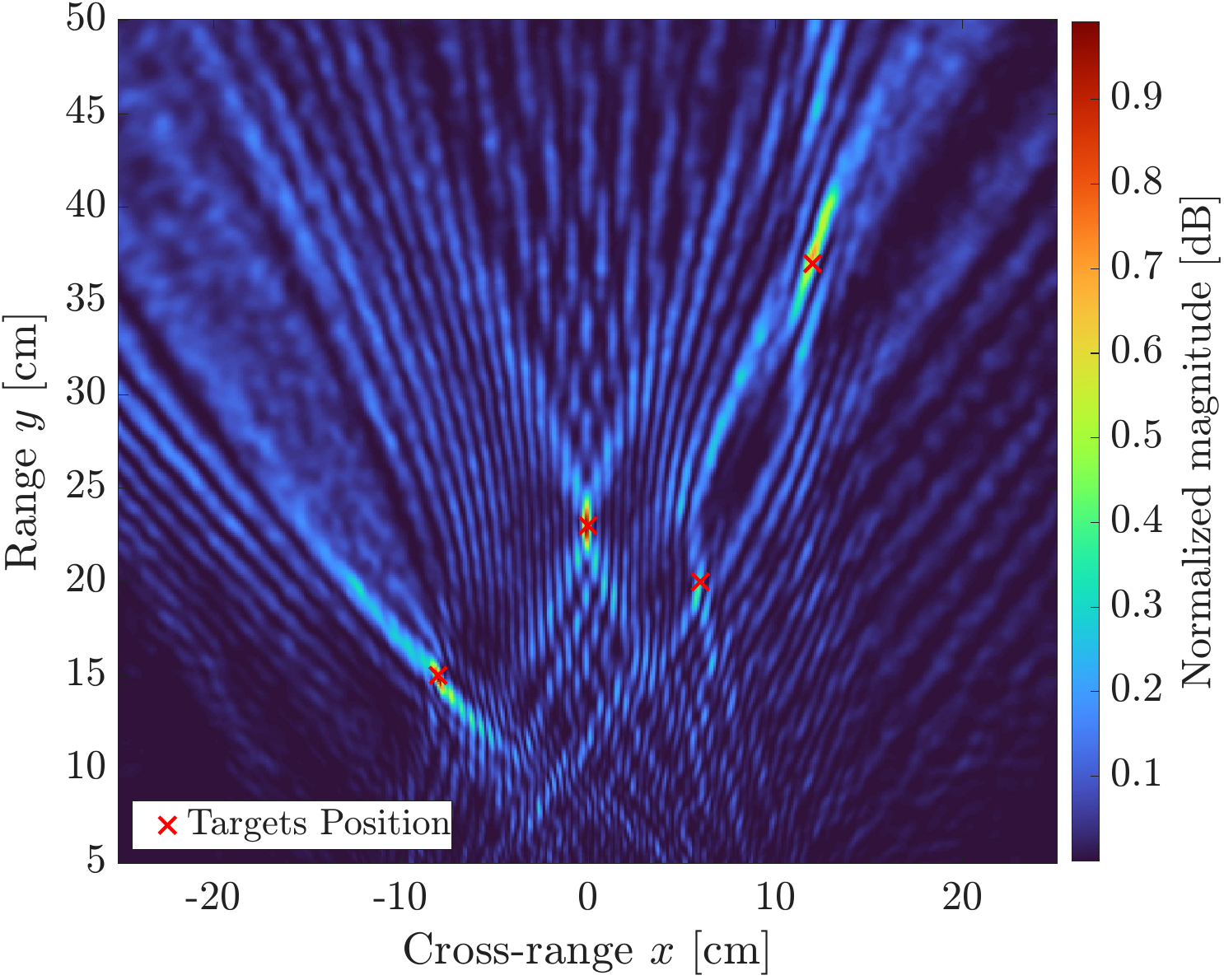}
            \caption{EFK-based Autofocus.}
            \label{fig:autofocus_iekf}
        \end{subfigure}
    \end{tabular}

    \caption{%
    \textbf{EKF-based autofocus: trajectory correction and imaging impact, with $\mathbf{SNR=10}$ dB.}
    \textbf{(a)} Platform trajectory in the $(x,y)$ plane, comparing the \emph{true} path, the \emph{IMU-derived} path affected by drift/noise, and the \emph{EKF-corrected} estimate. Targets are overlaid to show the scene geometry relative to the virtual aperture. 
    \textbf{(b)} \emph{Oracle} image obtained by \eqref{eq:bp} using the true trajectory, representing the best achievable coherent focusing; 
    \textbf{(c)} Image obtained by \eqref{eq:bp} using the raw IMU trajectory: trajectory errors introduce range-dependent phase mismatches across slow time, causing loss of coherent integration, defocusing, and target displacement/smearing.
    \textbf{(d)} Image obtained after autofocus using the proposed EKF method: the recovered path restores inter-pulse phase consistency.}
    \label{fig:iekf_autofocus_4panel}
\end{figure*}
\subsection{Bayesian Information Matrix}

The unknown parameters are the target position $\mathbf{p}\in\mathbb{R}^3$, the aperture errors
$\boldsymbol{\delta} = [\boldsymbol{\delta}_1^T,\ldots,\boldsymbol{\delta}_{M-1}^T]^T$ (with $\boldsymbol{\delta}_0=\mathbf{0}$), and the complex reflectivity $\alpha\in\mathbb{C}$. The likelihood is complex Gaussian with mean $\{\boldsymbol{\mu}_m\}$ and covariance $\sigma_w^2\mathbf{I}_K$.

For real parameters, the (data) Fisher information matrix has the standard form
\begin{equation}
[\mathbf{J}]_{ij}
=
\frac{2}{\sigma_w^2}\,
\Re\!\left\{
\sum_{m=0}^{M-1}
\left(\frac{\partial \boldsymbol{\mu}_m}{\partial \theta_i}\right)^{\!H}
\left(\frac{\partial \boldsymbol{\mu}_m}{\partial \theta_j}\right)
\right\},
\label{eq:fim_general}
\end{equation}
where $(\cdot)^H$ denotes the Hermitian transpose. The derivatives with respect to $\mathbf{p}$ and $\boldsymbol{\delta}$ follow by the chain rule through $r_m=\|\mathbf{p}-\mathbf{q}_m\|$ and $\mathbf{q}_m=\hat{\mathbf{q}}_m-\boldsymbol{\delta}_m$:
\begin{equation}
\frac{\partial \boldsymbol{\mu}_m}{\partial \theta}
=
\frac{\partial \boldsymbol{\mu}_m}{\partial r_m}\,
\frac{\partial r_m}{\partial \theta},
\qquad \theta\in\{\mathbf{p},\boldsymbol{\delta}\}.
\label{eq:chain_rule}
\end{equation}
Incorporating the Gaussian prior~\eqref{eq:delta_prior_imu} on $\boldsymbol{\delta}$ yields the Bayesian information matrix by augmenting the $\boldsymbol{\delta}\boldsymbol{\delta}$ block:
\begin{equation}
\boldsymbol{\Psi} \triangleq \mathbf{J}_{\boldsymbol{\delta}\boldsymbol{\delta}}+\mathbf{C}_{\boldsymbol{\delta}}^{-1}.
\label{eq:Psi_def}
\end{equation}
The BCRB for $\mathbf{p}$ follows from the Schur complement~\cite{VanTrees1968}:
\begin{equation}
\mathrm{BCRB}(\mathbf{p}) =
\left(
\mathbf{J}_{\mathbf{p}\mathbf{p}}
-
\mathbf{J}_{\mathbf{p}\boldsymbol{\delta}}\,
\boldsymbol{\Psi}^{-1}\,
\mathbf{J}_{\mathbf{p}\boldsymbol{\delta}}^{T}
-
\mathbf{G}\,\mathbf{S}_\alpha^{-1}\,\mathbf{G}^{T}
\right)^{-1},
\label{eq:bcrb_general}
\end{equation}
where
$\mathbf{S}_\alpha = \mathbf{J}_{\alpha\alpha} - \mathbf{J}_{\boldsymbol{\delta}\alpha}^{T}\boldsymbol{\Psi}^{-1}\mathbf{J}_{\boldsymbol{\delta}\alpha}$ and
$\mathbf{G}=\mathbf{J}_{\mathbf{p}\alpha}-\mathbf{J}_{\mathbf{p}\boldsymbol{\delta}}\boldsymbol{\Psi}^{-1}\mathbf{J}_{\boldsymbol{\delta}\alpha}$.
The three terms represent, respectively, information from the observations, loss due to aperture uncertainty, and loss due to the unknown reflectivity.

%==============================================================================
\section{Coherent Aperture Synthesis}
\label{sec:processing}
%==============================================================================

The following section develops an autofocus algorithm that estimates and compensates the trajectory errors, enabling near-bound performance. The procedure starts from the range-compressed profiles in~\eqref{eq:z_ell_sinc}, forms a near-field image by backprojection, and refines the aperture trajectory through an autofocus driven by calibration points extracted from an initial short-aperture image. The effect of trajectory correction on image quality is illustrated in Fig.~\ref{fig:iekf_autofocus_4panel}, where the defocusing caused by IMU drift and its compensation through the EKF are compared for a representative scenario.

\subsection{Backprojection Imaging}

Let $\mathcal{G}=\{\mathbf{r}_g\}_{g=1}^{G}$ denote a grid of candidate positions. For each aperture index $m$, antenna $n$, and pixel $\mathbf{r}_g$, define the predicted range $\hat{r}_{n,g,m} = \|\mathbf{r}_g - \hat{\mathbf{p}}_{n,m}\|$, where $\hat{\mathbf{p}}_{n,m} = \hat{\mathbf{q}}_m + \mathbf{d}_n$ is the estimated antenna position. The backprojection image is
\begin{equation}
I(\mathbf{r}_g)=\sum_{m=0}^{M-1}\sum_{n=1}^{N} z_{n,m}[\hat{\nu}_{n,g,m}]\,e^{j\kappa \hat{r}_{n,g,m}},
\label{eq:bp}
\end{equation}
where $\kappa = 4\pi/\lambda$, $\hat{\nu}_{n,g,m} = (2B/c)\hat{r}_{n,g,m}$, and $z_{n,m}[\nu]$ denotes fractional-delay interpolation of the range-compressed profile.

\subsection{Autofocus}

Trajectory errors $\boldsymbol{\delta}_m = \hat{\mathbf{q}}_m - \mathbf{q}_m$ perturb the predicted range in~\eqref{eq:bp}. For a scatterer at $\mathbf{r}_g$, the true range is $r_{n,g,m} = \|\mathbf{r}_g - \mathbf{p}_{n,m}\|$ while the predicted range uses the erroneous position $\hat{\mathbf{p}}_{n,m}$. First-order expansion yields
\begin{equation}
\hat{r}_{n,g,m} - r_{n,g,m} \approx -\mathbf{u}_{n,g,m}^T\boldsymbol{\delta}_m,
\label{eq:range_err}
\end{equation}
where $\mathbf{u}_{n,g,m} = (\mathbf{r}_g - \mathbf{p}_{n,m})/r_{n,g,m}$ is the unit look vector. The corresponding phase error is $\Delta\phi_{n,g,m} = \kappa(\hat{r}_{n,g,m} - r_{n,g,m}) \approx -\kappa\,\mathbf{u}_{n,g,m}^T\boldsymbol{\delta}_m$. Since $\boldsymbol{\delta}_m$ varies with $m$ and the look vector $\mathbf{u}_{n,g,m}$ varies with both $m$ and target position, the phase error is space-variant and cannot be removed by a single correction applied uniformly across the image.

We estimate $\boldsymbol{\delta}_m$ using phase observations from a set of calibration points whose positions are known. A provisional image $I_0(\mathbf{r}_g)$ is formed from the first $M_0 \ll M$ acquisitions using~\eqref{eq:bp}. Since IMU errors grow gradually with aperture length, this short-aperture image remains focused and permits extraction of a calibration set $\mathcal{C} = \{\mathbf{c}_q\}_{q=1}^{Q}$ as the $Q$ strongest local maxima of $|I_0|^2$. These points are assumed to be stationary over the full aperture.

For acquisition $m$, antenna $n$, and calibration point $\mathbf{c}_q$, the carrier phase is extracted by matched filtering at the predicted delay $\hat{\tau}_{n,q,m} = 2\hat{r}_{n,q,m}/c$. The matched filter output is
\begin{equation}
\alpha_{n,q,m} = \sum_{\ell} z_{n,m}[\ell]\, S^*(\ell - \hat{\nu}_{n,q,m}),
\label{eq:mf_output}
\end{equation}
where $S(\cdot)$ is the Dirichlet kernel and $\hat{\nu}_{n,q,m} = (2B/c)\hat{r}_{n,q,m}$. The extracted phase is $\phi_{n,q,m} = \arg(\alpha_{n,q,m})$.

To eliminate absolute phase ambiguity arising from unknown scatterer reflectivity, differential measurements are formed relative to the first acquisition:
\begin{equation}
\psi_{n,q,m} = \phi_{n,q,m} - \phi_{n,q,0}.
\label{eq:diff_phase}
\end{equation}
This differencing cancels the unknown phase contribution from the scatterer while preserving sensitivity to trajectory errors, since $\boldsymbol{\delta}_0 = \mathbf{0}$ by construction.

The predicted differential phase, given trajectory error $\boldsymbol{\delta}_m$, follows from the round-trip propagation:
\begin{equation}
g_{n,q,m}(\boldsymbol{\delta}_m) = -\kappa\bigl(r_{n,q,m}(\boldsymbol{\delta}_m) - r_{n,q,0}\bigr),
\label{eq:g_model}
\end{equation}
where $r_{n,q,m}(\boldsymbol{\delta}_m) = \|\mathbf{c}_q - (\hat{\mathbf{p}}_{n,m} - \boldsymbol{\delta}_m)\|$ is the corrected range. Linearization of~\eqref{eq:g_model} about $\boldsymbol{\delta}_m = \mathbf{0}$ yields the Jacobian
\begin{equation}
\frac{\partial g_{n,q,m}}{\partial \boldsymbol{\delta}_m} = \kappa\,\mathbf{u}_{n,q,m}^T,
\label{eq:jacobian}
\end{equation}
where $\mathbf{u}_{n,q,m} = (\mathbf{c}_q - \hat{\mathbf{p}}_{n,m})/\hat{r}_{n,q,m}$. Observability requires that the $NQ$ look vectors at each acquisition span the position error subspace; this is ensured when calibration points are distributed in angle around the array.

Let $\boldsymbol{\xi}_m$ denote the state vector comprising $\boldsymbol{\delta}_m$ and auxiliary variables from the inertial error model. The state evolves according to
\begin{equation}
\boldsymbol{\xi}_{m+1} = F\boldsymbol{\xi}_m + \mathbf{w}_m,
\label{eq:state_transition}
\end{equation}
where $F$ and $\mathrm{Cov}(\mathbf{w}_m) = Q$ are determined by the error dynamics. At each acquisition $m \geq 1$, the $NQ$ differential phase measurements are stacked into $\boldsymbol{\psi}_m \in \mathbb{R}^{NQ}$. These are incorporated sequentially through an extended Kalman filter~\cite{Gelb1974}. The linearized observation model is
\begin{equation}
\boldsymbol{\psi}_m = G_m \boldsymbol{\xi}_m + \boldsymbol{\eta}_m,
\label{eq:obs_model}
\end{equation}
where the rows of $G_m$ contain the Jacobians from~\eqref{eq:jacobian} and $\boldsymbol{\eta}_m \sim \mathcal{N}(\mathbf{0}, \Sigma_\psi)$ represents phase extraction noise with covariance $\Sigma_\psi = \sigma_\phi^2 I_{NQ}$. The filter recursion yields the minimum-variance estimate $\hat{\boldsymbol{\xi}}_m$ at each acquisition, from which the position error estimate $\hat{\boldsymbol{\delta}}_m$ is extracted. The corrected trajectory is
\begin{equation}
\mathbf{q}_m^{\mathrm{corr}} = \hat{\mathbf{q}}_m - \hat{\boldsymbol{\delta}}_m.
\label{eq:corrected_traj}
\end{equation}

%==============================================================================
\section{Numerical Results}
\label{sec:results}
%==============================================================================
In this section, we illustrate (i) the accuracy limits of opportunistic aperture sensing under inertial uncertainty, and (ii) the resulting transmit-power headroom enabled by distance-aware exposure control. 

\begin{figure}[t!]
    \centering
    \includegraphics[width=0.9\columnwidth]{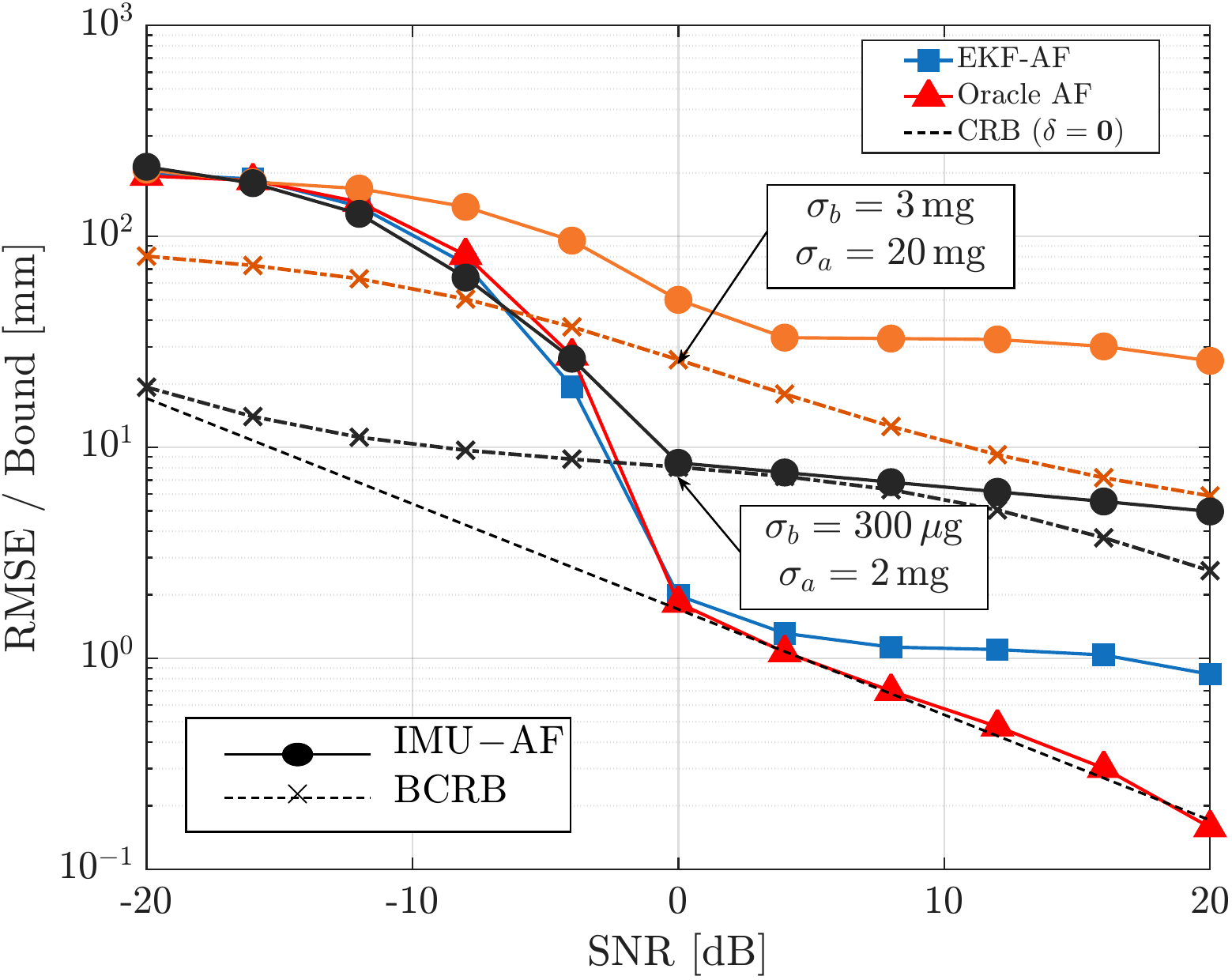}
    \caption{Localization RMSE versus SNR together with the ideal CRB (known aperture) and the BCRB (correlated inertial uncertainty). Oracle-AF: localization using known trajectory, EKF-AF: localization using the EKF-corrected trajectory, and IMU-AF: localization using the raw IMU trajectory.}
    \label{fig:rmse_vs_snr}
\end{figure}

We consider a 28~GHz device equipped with a compact $N=4$ element array. The device performs monostatic sensing using the uplink waveform, forming a virtual aperture of length $A$ over $M=\lceil 4A/\lambda\rceil$ slow-time acquisitions.
Inertial errors follow \eqref{eq:Ct}. Two representative IMU quality points are reported in the bound-validation plot: a \emph{consumer-grade} case (high bias and noise) and a \emph{high-grade} case (low bias and noise), matching the parameter pairs annotated in Fig.~\ref{fig:rmse_vs_snr}.

\begin{table}[b]
\centering
\caption{Parameters used in the numerical evaluation.}
\label{tab:sim_params_results}
\begin{tabular}{lcc}
\hline
\textbf{Quantity} & \textbf{Symbol} & \textbf{Value} \\
\hline
Carrier frequency & $f_c$ & 28~GHz \\
Bandwidth & $B$ & 200~MHz \\
Power-density limit & $S_{\lim}$ & 10~W/m$^2$ \\
FCC averaging distance (baseline) & $r_s$ & 2.5~cm \\

\hline
\end{tabular}
\end{table}

%------------------------------------------------------------------------------
%------------------------------------------------------------------------------
\subsection{Validation of CRB/BCRB and of EKF-based Autofocus}
\label{sec:results_bounds}
%------------------------------------------------------------------------------
Fig.~\ref{fig:rmse_vs_snr} quantifies the localization accuracy as a function of SNR for two representative IMU configurations. The dashed curve labeled $\mathrm{CRB}(\boldsymbol{\delta}=0)$ corresponds to ideal coherent synthesis with a perfectly known aperture and decreases monotonically with SNR. The $\mathrm{BCRB}$ curves, computed from~\eqref{eq:bcrb_general}, account for the correlated trajectory uncertainty induced by inertial sensing and exhibit a floor at high SNR where the position prior dominates the Fisher information.

The curves labeled IMU-AF and EKF-AF report the empirical RMSE obtained by backprojection using, respectively, the raw IMU trajectory and the EKF-corrected trajectory. Three observations are in order. First, the IMU-AF performance degrades rapidly as SNR increases beyond the point where trajectory errors become the dominant source of localization uncertainty; the RMSE saturates at a level consistent with the BCRB floor. Second, the EKF-based autofocus substantially reduces the effective trajectory error: at low-to-moderate SNR, the EKF-AF curve closely tracks the ideal CRB, indicating that the phase observations from calibration points provide sufficient information to compensate the inertial drift. Third, the EKF-AF curve itself exhibits a floor at high SNR, approximately one order of magnitude above the CRB. This residual error arises from uncompensated phase estimation noise in the autofocus loop and represents a practical limitation of the proposed algorithm. The floor level depends on the number and geometric distribution of calibration points, as well as on the phase extraction SNR; its reduction constitutes a direction for further investigation.

%------------------------------------------------------------------------------
\subsection{Exposure-Control Gain: EIRP Enabled by Distance Awareness}
\label{sec:results_eirp}
%------------------------------------------------------------------------------
We now translate localization uncertainty into a conservative exposure-compliant EIRP.
In this paper we adopt the FCC MPE limit above $6~\mathrm{GHz}$,
denoted by $S_{\lim}$ in $\mathrm{W/m^2}$.
Let $r$ be the separation between the device and the head/torso region of interest.
In the far-field regime, compliance with $S(r)\le S_{\lim}$ is ensured by the inverse-square bound
\begin{equation}
\mathrm{EIRP} \;\le\; 4\pi S_{\lim} r^2,
\label{eq:eirp_limit_ff}
\end{equation}
where $\mathrm{EIRP}$ is the equivalent isotropically radiated power.
Commercial devices implement exposure mitigation via proximity sensing, which indicates whether a
conductive body is within a short range but does not provide a reliable metric distance
\cite{FCC_KDB447498}.
Accordingly, when the device is recognized as \emph{handheld}, the \emph{baseline} must guarantee
compliance under the worst case, i.e., $r=r_s$, where $r_s$ denotes the effective triggering distance
used for certification and control.
The resulting conservative EIRP budget is therefore $\mathrm{EIRP}_{\mathrm{base}} \;=\; 4\pi S_{\lim} r_s^2$,
which is enforced throughout the on-body/handheld regimes.
In addition, the device may be certified to radiate a larger maximum power in an \emph{off-body} state.
Denote by $\mathrm{EIRP}_{\max}$ the maximum radiated power allowed by the radio chain.
We assume the handset can classify its operational state as:
(i) \emph{on-body} for $r\le r_s$,
(ii) \emph{handheld} for $r_s<r\le r_{\mathrm{off}}$,
and (iii) \emph{off-body} for $r>r_{\mathrm{off}}$, with $r_{\mathrm{off}}=50~\mathrm{cm}$.
The baseline policy is thus
\begin{equation}
\mathrm{EIRP}_{\mathrm{base}}(r)=
\begin{cases}
\mathrm{EIRP}_{\mathrm{base}}, & r \le r_{\mathrm{off}},\\[2pt]
\mathrm{EIRP}_{\max}, & r> r_{\mathrm{off}}.
\end{cases}
\label{eq:baseline_policy_states}
\end{equation}

\begin{figure}[b!]\vspace{-0.5cm}
    \centering
    \includegraphics[width=0.9\columnwidth]{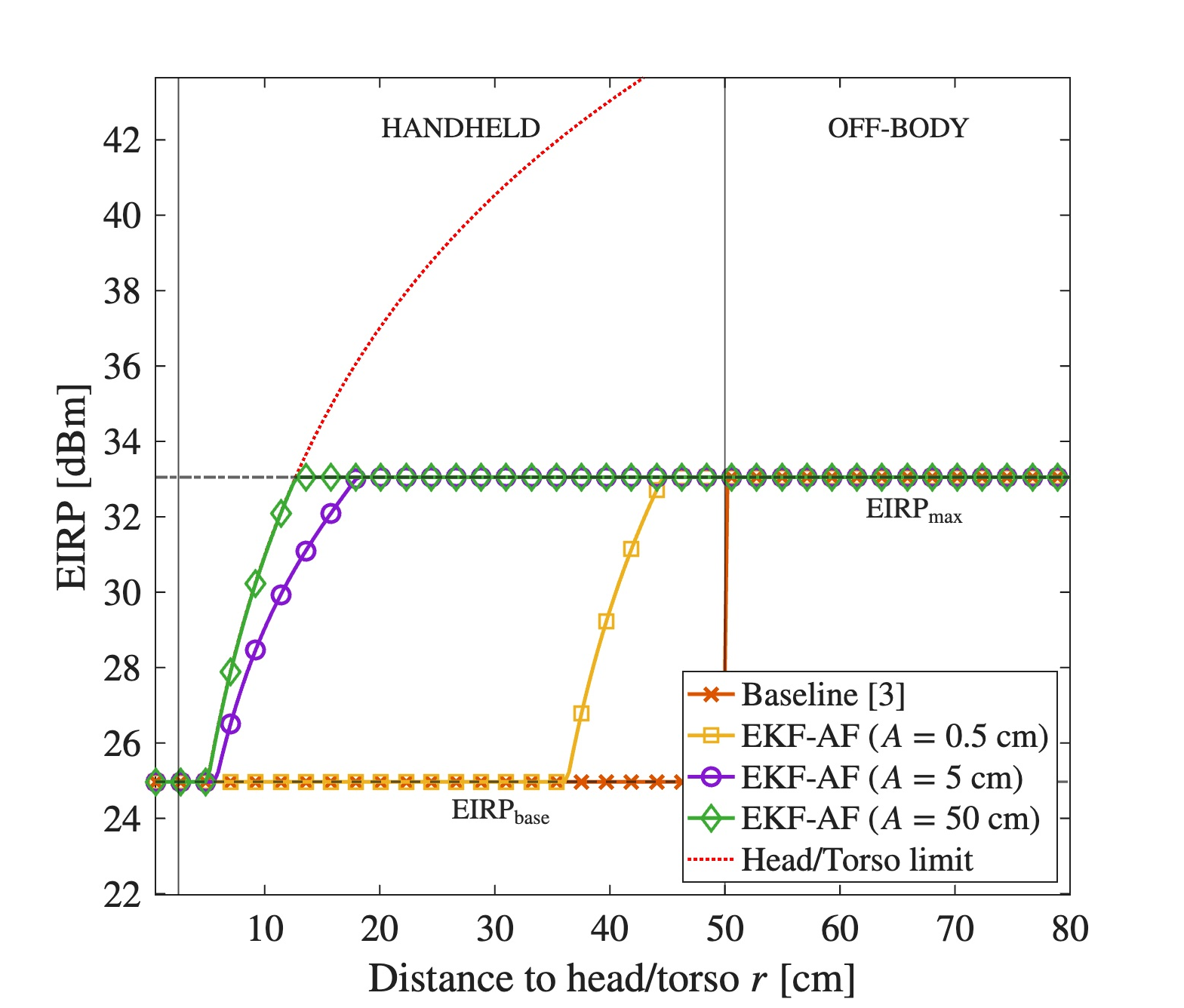}
    \caption{Maximum allowable EIRP versus head/torso distance under FCC power-density constraint. Baseline: binary proximity-driven compliance at $r_s=2.5$~cm. Proposed: distance-aware policy using $r_{\mathrm{eff}}=\hat r-k\sqrt{\mathrm{CRB}(\hat r)}$ with $k=2.58$, saturating at $\mathrm{EIRP}_{\max}$ for off-body operation ($r>50$~cm).}
    \label{fig:eirp_vs_r}
\end{figure}

The proposed method exploits device-centric ISAC to estimate the head/torso distance $\hat r$ and
associates to it a guard margin derived from the validated bound in
Section~\ref{sec:crb}.
Specifically, we define the effective distance
\begin{equation}
r_{\mathrm{eff}} \triangleq \hat r - k \sqrt{\mathrm{CRB}(\hat r)},
\qquad
\Pr\{r \ge r_{\mathrm{eff}}\}\approx 0.99,
\label{eq:reff_results}
\end{equation}
where $k=2.58$ yields a conservative one-sided coverage under a Gaussian error model.
The exposure-compliant EIRP is then chosen as
\begin{equation}
\mathrm{EIRP}_{\mathrm{prop}}(\hat r)
=
\min\!\Big\{4\pi S_{\lim} r_{\mathrm{eff}}^{\,2},\ \mathrm{EIRP}_{\max}\Big\}.
\label{eq:eirp_prop_results}
\end{equation}
Hence, distance awareness converts the binary back-off into a distance-proportional policy in the
handheld regime: the admissible EIRP increases with separation until it saturates at
$\mathrm{EIRP}_{\max}$ for $r>r_{\mathrm{off}}$. 
Fig.~\ref{fig:eirp_vs_r} compares the admissible EIRP as a function of head/torso separation for the baseline proximity-based policy and the proposed distance-aware policy. The baseline, indicated by the horizontal segment at $\mathrm{EIRP}_{\mathrm{base}} \approx 25$~dBm, enforces a constant power limit throughout the handheld regime regardless of the actual separation. The red dotted curve represents the physical MPE constraint from~\eqref{eq:eirp_limit_ff}, which increases quadratically with distance.

The three curves labeled EKF-AF correspond to the proposed policy~\eqref{eq:eirp_prop_results} evaluated at SNR~$=$~5~dB for aperture lengths $A \in \{0.5,\, 5,\, 50\}$~cm. The dependence on aperture length admits a direct physical interpretation: a longer virtual aperture provides finer angular resolution, which translates into a smaller localization variance and hence a smaller guard margin $k\sqrt{\mathrm{CRB}}$. Consequently, the effective distance $r_{\mathrm{eff}}$ approaches the measured distance $\hat{r}$ more closely, and the admissible EIRP increases.

For the longest aperture ($A = 50$~cm), the proposed policy reaches $\mathrm{EIRP}_{\max}$ at approximately $r = 15$~cm, yielding a gain of roughly 8~dB over the baseline in the range $10 \le r \le 20$~cm. At shorter apertures the transition is more gradual: the $A = 5$~cm curve attains $\mathrm{EIRP}_{\max}$ near $r = 20$~cm, while the $A = 0.5$~cm curve remains below $\mathrm{EIRP}_{\max}$ until $r \approx 45$~cm. These results indicate that even modest hand motion during a transmission interval can provide sufficient aperture diversity to unlock significant EIRP headroom in the handheld regime.

%==============================================================================
%==============================================================================
\section{Conclusion}
\label{sec:conclusion}
%==============================================================================

This paper has developed a device-centric integrated sensing and communication framework that enables handheld devices to measure their distance from the user's body and adjust transmit power accordingly. The proposed method exploits the natural motion of the user's hand to synthesize a virtual sensing aperture, using the uplink communication waveform itself as the probing signal. Because the aperture trajectory is not known precisely, an extended Kalman filter was employed to estimate and compensate the position errors introduced by inertial sensors.

The theoretical analysis established the Bayesian Cram\'{e}r--Rao bound for target localization under correlated aperture position uncertainty, revealing that the achievable accuracy is ultimately limited by the quality of the inertial measurements when autofocus is not applied.

Numerical results at 28~GHz demonstrated that the EKF-based autofocus closely approaches the ideal CRB at low-to-moderate SNR, achieving centimeter-level localization accuracy with realistic motion parameters. At high SNR, a residual floor appears due to uncompensated phase noise in the autofocus loop; reducing this floor through improved calibration strategies remains an open problem. When the localization accuracy is translated into exposure-compliant EIRP, the distance-aware policy provides up to 8~dB of additional power headroom compared to the binary proximity-based baseline, with longer apertures yielding larger gains due to improved angular resolution.

%==============================================================================

%==============================================================================
% REFERENCES
%==============================================================================

\bibliographystyle{IEEEtran}
\bibliography{references}

\end{document}